\documentclass[a4paper]{article}
\usepackage[numbers]{natbib}

\usepackage{template/INTERSPEECH2021}
\title{VoiceMe: Personalized voice generation in TTS}
\name{Pol van Rijn$^1$\thanks{Audio and video samples and the supplementary figure can be found here: https://polvanrijn.github.io/VoiceMe/ All code is here: https://github.com/polvanrijn/VoiceMe}, Silvan Mertes$^2$, Dominik Schiller$^2$, Piotr Dura$^3$, Hubert Siuzdak$^3$, Peter M. C. Harrison$^4$, Elisabeth André$^2$, Nori Jacoby$^1$}
\address{
  $^1$Max-Planck-Institute for Empirical Aesthetics, Frankfurt, Germany\\
  $^2$Human-Centered Artificial Intelligence, Augsburg, Germany\\
  $^3$Charactr Inc., Los Angeles, US\\
  $^4$University of Cambridge, Cambridge, UK
}
\email{pol.van-rijn@ae.mpg.de}

\begin{document}

\maketitle

\begin{abstract}
Novel text-to-speech systems can generate entirely new voices that were not seen during training. However, it remains a difficult task to efficiently create personalized voices from a high-dimensional speaker space. In this work, we use speaker embeddings from a state-of-the-art speaker verification model (SpeakerNet)  trained on thousands of speakers to condition a TTS model. We employ a human sampling paradigm to explore this speaker latent space. We show that users can create voices that fit well to photos of faces, art portraits, and cartoons. We recruit online participants to collectively manipulate the voice of a speaking face. We show that  (1) a separate group of human raters confirms that the created voices match the faces, (2) speaker gender apparent from the face is well-recovered in the voice, and (3) people are consistently moving towards the real voice prototype for the given face. Our results demonstrate that this technology can be applied in a wide number of applications including character voice development in audiobooks and games, personalized speech assistants, and individual voices for people with speech-impairment.
\end{abstract}

\section{Introduction}
In the last few years, multi-speaker text-to-speech (TTS) models have been proposed that can create entirely new high-quality voices  \cite{jia2018transfer, stanton2021speaker}. While these approaches can generate unique voices that are distinct from voices seen during training, it is unclear how to create a personalized voice that fits one's own mental representation. The existence of such a tool would open the door to numerous creative and practical applications, such as developing customized voices for robots, personalized speech assistants, bringing fictional characters or paintings to life, or developing individualized voices for speech-impaired people. 

A large body of psychological research has shown that people actively make inferences about faces, including the personality, age, or background of the person \cite{jack2015review, sofer2014trustworthiness, ekman1992emotions, angulu2018age, walker2016personality}. Since the relation between face and voice is complex, we use humans-in-the-loop to find voices in the speaker latent space of a trained TTS model that match their perception of faces in photographs, art portraits, and cartoons. Despite the fact that this approach is not automated and relies on expensive human judgments, subjective evaluations from an independent group of human raters confirm that the created voices accurately match the faces.

\section{Background}
Previous human-in-the-loop approaches have shown that humans can iteratively build feature representations for personalizing sound and speech. For example, Ritschel et al. \cite{ritschel2019personalized} used an interactive evolutionary algorithm to allow users to create sounds that express intentions and emotions for a social robot. Van Rijn et al. \cite{vanrijn2021exploring} used an adaptive method to sample from latent semantic human representations \cite{harrison2020gsp} to find prototypes of emotional speech by adapting the latent representation of a GST Tacotron model \cite{wang2018style}.

In the past years, various approaches have been proposed to create novel voices. Jia et al. \cite{jia2018transfer} demonstrated that an independently trained speaker encoder network trained on a speaker verification task can produce useful conditioning for a multi-speaker text-to-speech model. By sampling random points from the obtained speaker embedding space, the authors generated fictitious voices that were not seen during the training. Another approach was proposed by Stanton et al. \cite{stanton2021speaker} that does not rely on transfer learning from the speaker verification task, but jointly learns a distribution over speaker embeddings, also allowing for sampling a novel voice. However, all of these papers focus on speaker generation, but not on speaker personalization. To our knowledge, this is the first paper using speaker generation models for voice personalization. 

\begin{figure*}[ht!]
    \centering
    \includegraphics{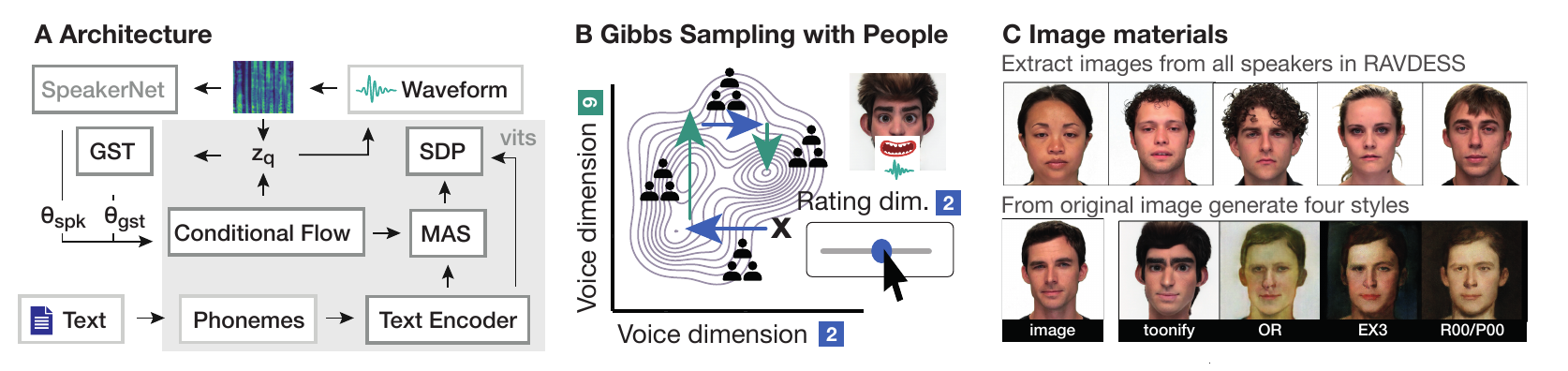}
    \caption{Overview. (A) The architecture used in this paper is a modified version of the VITS model using SpeakerNet ($\theta_\mathrm{spk}$) and GST embeddings ($\theta_\mathrm{gst}$). $z_q$ is the posterior latent sequence for a speech sample. (B) Participants move a slider adjusting a single dimension of the voice. The same slider is presented to three participants. The mean answer is passed to the next iteration. This allows us to gradually optimize the voice which fits to a face over the course of iterations. (C) Image materials. Images are extracted for all speakers in the RAVDESS corpus. We use deep-learning style transfer to convert the images to cartoons and paintings.}
    \label{fig1}
\end{figure*}

In the present paper, we adopt the method proposed by Jia et al. \cite{jia2018transfer} and used embeddings from a speaker verification network as speaker representation: because they can be trained on noisy speech of thousands of speakers, do not require transcripts, can extract speaker embeddings for unseen voices, and the obtained voice prototypes can be reused in future models if trained on the same pretrained speaker verification network. While this work use this particular state-of-the-art architecture for voice synthesis (Figure \ref{fig1}A), our human-sampling approach (Figure \ref{fig1}B) can be extended to work with a large class of speaker generation models so that they can be used for voice personalization.

\section{Methods}
\subsection{TTS Architecture}
Here, we use the state-of-the-art TTS model VITS \cite{kim2021conditional}, consisting of the following components: (i) text frontend composed of text normalization followed by a grapheme-to-phoneme model utilizing IPA characters (International Phonetic Alphabet) \cite{bernard2021phonemizer}, (ii) transformer-based Text Encoder with a projection layer used to construct prior distribution, (iii) Normalizing Flow greatly improving the flexibility of prior latent space, (iv) convolutional Posterior Encoder producing $z_q$, (v) Monotonic Alignment Search (MAS) and Stochastic Duration Predictor (SDP) modules learning to align input characters to encoded spectrogram frames, (vi) high-fidelity GAN vocoder based on the HiFi-GAN \cite{jungil2020hifigan} architecture. We extend VITS by two additional components described below: SpeakerNet and GST.

We use the pretrained speaker verification network SpeakerNet-M \cite{koluguri2020speakernet} as speaker embeddings. The model was trained on 7,205 speakers and has a satisfactory trade-off between the quality on a speaker recognition task and a small computational footprint. We furthermore found that SpeakerNet mainly encoded voice pitch and timbre information and not prosody. In order to learn prosodic variation, we extend our model with a bank of Global Style Tokens (GST) \cite{wang2018style}, which extracts style embeddings from encoded spectrogram frames. We initialized 16 Global Style Tokens with 8 attention heads and the resulting embedding size was set to 256. For the GST encoder, an 8-layer convolutional network was used with the same architecture as the Posterior Encoder. During training, the speaker and style embedding were separately L2-normalized and concatenated. During the experiments, we use the same zero embedding to keep prosody approximately constant across samples.

To prevent the GSTs from learning speaker-dependent features in presence of SpeakerNet embeddings, an additional adversarial loss is proposed as follows. Alongside discriminators, a separate shallow feed-forward neural network is trained to reconstruct the speaker embeddings from extracted style embeddings. During the discriminator step, this network minimizes a cosine distance between real and reconstructed speaker embeddings using cosine embedding loss: $(1-\mathrm{cos}(x,\hat{x}))$. Conversely, during the generator step, the style extractor is penalized if this network succeeded in reconstructing speaker embeddings -- the loss function is then: $(\mathrm{max}(0,\mathrm{cos}(x,\hat{x})))$. The overview of the architecture is depicted in Figure \ref{fig1}A.

We applied transfer learning from the publicly available VCTK checkpoint \cite{vits} and the training was continued using two NVIDIA V100 GPUs. For the first 400k iterations only the discriminators were allowed to train due to the lack of a published discriminator checkpoint, then normal training continued for additional 2M iterations with learning rate lowered to 1e-4.

\begin{figure*}[ht!]
    \centering
    \includegraphics{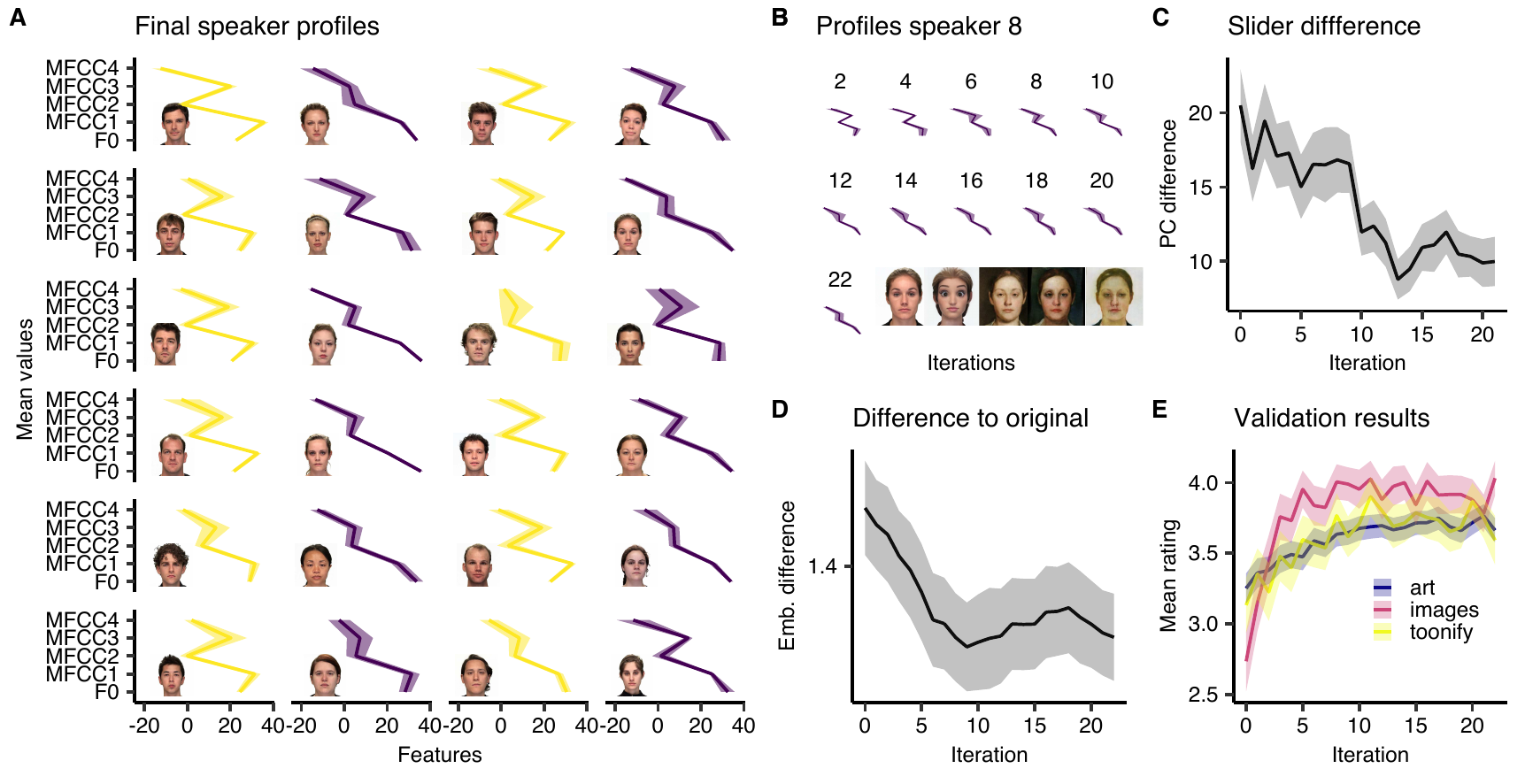}
    \caption{Results. (A) Final speaker profiles on mean values for F0 and MFCC1--4. The shaded areas in all plots refer to 95\% confidence intervals. (B) Example profiles for speaker 8 at different iterations. (C) The Euclidean distance between consecutive iterations is larger for earlier iterations compared to later iterations. (D) The Euclidean distance to the original reference drops over the first ten dimensions and slightly increases and decreases. (E) The mean opinion score increases over the course of iterations.
    }
    \label{fig2}
\end{figure*}
\subsection{Gibbs Sampling with People}
`Gibbs Sampling with People' (GSP) is an adaptive technique \cite{harrison2020gsp}, 
where many participants collaborate to navigate a stimulus space and identify regions associated with a given semantic concept, in our case optimize voice parameters so that they match a target face animation.  The participants' responses are organized into sequences of iterations called ``chains''.  In each trial, the participant is presented with a stimulus (a synthesized voice) and a slider, where the slider is coupled to a particular dimension of the stimulus space that changes from trial to trial. The participant is instructed to move the slider to find the stimulus most associated with the target face. In our implementation, three different participants contribute trials for a given iteration in a given chain, and their responses are averaged. The resulting stimulus (based on the average response) is then passed along the chain of participants, with each successive generation of participants optimizing a different dimension of the stimulus (Figure 1B). This procedure is repeated multiple times, cycling through each of the dimensions of the sample space. Harrison et al. (\cite{harrison2020gsp}) demonstrated that the emergent process corresponds to a Gibbs sampler that maps the relationship between the target face and the participants' internalized representations of the corresponding voice parameters

In the current experiment, participants change the first ten principal components of the SpeakerNet embeddings. Initial piloting suggested that these principal components had the desired property of intuitive interpretability (e.g., PC2 has a strong gender effect), and prior research with related models suggested that 10 principal components should be enough to achieve meaningful control over the stimuli \cite{harrison2020gsp}. The principal components were computed on SpeakerNet embeddings extracted on a single utterance of the 45,825 speakers present in the train, dev, and test partitions of the English CommonVoice dataset \cite{commonvoice} and account for 25.4 \% of the variance. The participants are prompted to adjust a slider that corresponds to one principal component to make the voice maximally similar to a face (see Figure \ref{fig1}C for some of the faces). For practical reasons, every slider contains a finite resolution of 31 equally-spaced slider positions. As opposed to using static images, we use Wav2Lip \cite{prajwal2020wav2lip} to synchronize the lips to the voice so that the resulting stimulus looks more natural (see supplementary materials).

\section{Materials}
\subsection{Stimuli}
In order to compare the personalized voice with ground truth, we extracted stills and speaker embeddings from recordings of real speakers. It was important to use speakers who are unknown to the participants because this might constrain possible voices attributed to the face. We used the same neutral utterance of all 24 speakers in the RAVDESS corpus \cite{livingstone2018ravdess}.

To demonstrate our approach does not only work for real faces, but also for fictional characters, we created for each original image four fictional characters based on style transfer. We used toonify \cite{pinkey2020toonify} and three additional art portrait styles from Ai Gahaku\cite{ai_gahaku}: OR, EX3, and ROO or P00 (see example in figure 1C). We selected the images in the following way.  We start by creating 12 art portraits and one toonified version and then select four styles with the highest perceptual similarity to the real photo \cite{zhang2018perceptual} (see Figure \ref{fig1}C). Thus, we select toonify, OR, and EX3 styles, but in 22 of 24 cases we select R00 and in all other cases we select P00.

For all 24 speakers, we use the extracted images and four styles with the highest perceptual similarity totaling 120 chains. To each chain, we randomly assign one of the 720 phonetically balanced and semantically neutral Harvard sentences \cite{harvardsentences}.

\subsection{Participants}
All participants were recruited from Amazon Mechanical Turk (AMT) and provided informed consent in accordance with the Max Planck Society Ethics Council approved protocol (application 2021\_42).  Participants were paid \$9/hour. Requirements for participation include a minimum age of 18 years, 99\% or higher approval rate on at least 5,000 previous tasks on AMT, residency in the US, and wearing headphones \cite{woods2017headphone}. Participant recruitment was managed by PsyNet \cite{harrison2020gsp}, an under-development framework for implementing complex experimental paradigms. This framework builds on the Dallinger platform \cite{dallinger} for experiment hosting and deployment.

\begin{figure*}[ht!]
    \centering
    \includegraphics{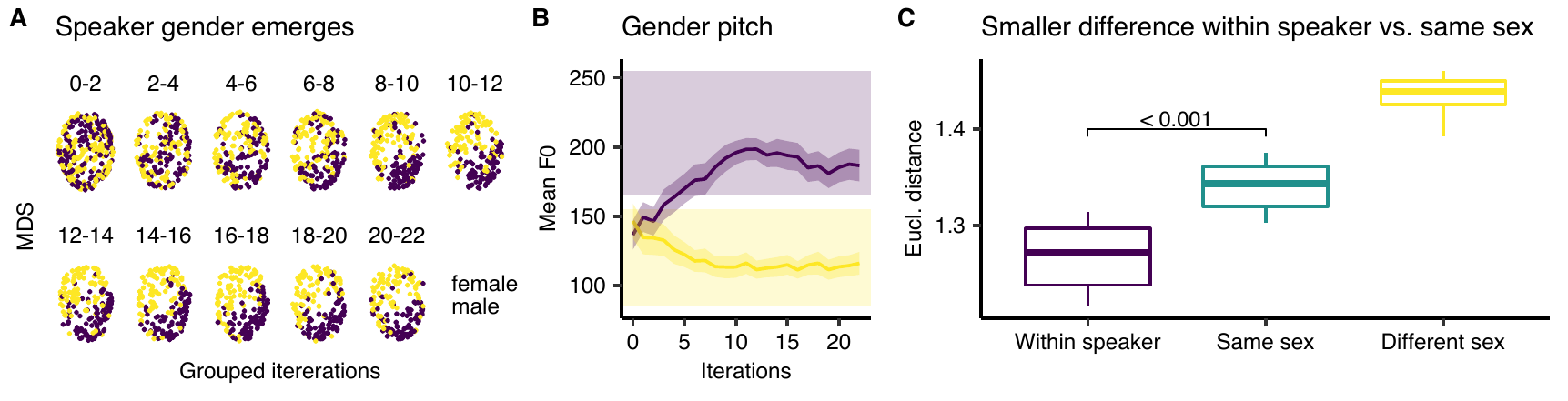}
    \caption{Gender difference. (A) Generated speaker embeddings in MDS space in which voices for male and female pictures occupy increasingly distinct areas in the voice latent space. Each row represents two iterations, read from left to right, top to bottom. (B) The pitch strongly increases for female faces and is lowered for male faces in the first ten iterations. The solid background reflects common pitch ranges for the sexes. Extracted F0 is expressed in Hz. (C) In the final iterations, the difference in the generated voice is smaller within the different styles of the same face (purple), compared to a random same-sex (green) or different-sex face (yellow).}
    \label{fig3}
\end{figure*}

\section{Results and discussion}
\subsection{Main experiment}
180 US participants (69 female, 3 prefer not to say, 68 male) engaged in the main experiment. The age ranged from 19 to 78 years old (\textit{M} = 41, \textit{SD} = 12). We terminated the experiment after 48 hours, after which 99 out of the 120 chains were full (22 iterations). In the last iteration, we obtain unique profiles for all 24 speakers (Figure \ref{fig2}A) and while each voice is randomly initialized, the profiles rapidly develop towards the final prototypes (Figure \ref{fig2}B). Quantitatively, we show that the Euclidean distance between consecutive iterations within a chain decreases over the course of iterations (Figure \ref{fig2}C), stabilizing toward the final 15 iterations. This means that participants move the sliders to a lesser extent at later iterations, suggesting convergence. We also compared the created speaker embedding at each iteration with the original speaker embedding of the real speaker. As we can see in Figure \ref{fig2}D, the difference to the reference is dropping for the first 10 iterations and then mildly increases and decreases again.

\subsection{Validation}
In a separate validation experiment, participants (\textit{N} = 110, 50 female, 1 prefer not to say, 59 male; age \textit{M} = 35, \textit{SD} = 10) rated how well the voice matches the moving face on a 5-point Mean Opinion Score (MOS): `Excellent', `Good', `Fair', `Poor', and `Bad match'. The validation included all stimuli generated in the first experiment, (overall 2,409 stimuli). Participants performed 200 ratings per experiment and consequently on average every stimulus was rated 9.1 times. As depicted in Figure \ref{fig2}E, the average MOS increases over the course of iterations for all styles. However, the increase is largest for the original faces moving from a 2.7 MOS at iteration 0 to a MOS of 4.0 (`Good match') in the later iterations (Wilcoxon rank sum test, \textit{Z} = .42, \textit{p} $<$ 0.001, Bonferroni-adjusted). The trend is followed by the cartoons (Wilcoxon rank sum test, \textit{Z} = .18, \textit{p} $<$ 0.001, Bonferroni-adjusted). 
For art portraits the improvement over iterations is smallest (Wilcoxon rank sum test, \textit{Z} = .16, \textit{p} $<$ 0.001, Bonferroni-adjusted).

\subsection{Toward personalized voice characteristics}
To further understand what kind of voice features were selected by the personalization process, we visualize the speaker latent space using Multi Dimensional Scaling on all voices created in the experiment. As shown in Figure \ref{fig3}A, over the course of iterations, male and female faces occupy increasingly distinct areas in the voice latent space. Furthermore, the average pitch starts at roughly the same point due to the random initialization of the voices and over the course of iterations is lowered for male and increased for female voices (Figure \ref{fig3}B). Voices for males and females converge in a pitch range common for the sex (85-155 and 165-255 Hz respectively) as indicated by the shaded areas \cite{fitch1970fundamental}.

Based on these results, we can state that the speaker gender apparent from the face is well-recovered in the voice. However, do people only focus on gender or also on other characteristics of the face? In order to address this question, we run another analysis. Here, we compute the Euclidean difference between the voices created for different styles of the same speaker versus a random speaker of the same sex. Using bootstrapping (\textit{n} = 1,000) we show that the voice differences within the same speaker are significantly smaller compared to a voice of a random speaker of the same sex (Figure \ref{fig3}C). The results show that the voice prototypes can capture face-specific image features in addition to gender.

\subsection{Limitations and outlook}
Since the ratings plateau at a MOS of 4 (``good match'') there is still room for improving the match. A simple way to improve the model is to increase the dimension of the latent space and apply GSP to the style embeddings instead of fixing the prosody in all chains. Furthermore, applying neutral prosody to all stimuli can have a negative impact on intelligibility, amplified by the fact that participants should focus on the match between voice and face rather than intelligibility. For future work, we may also explore running iterations within the same participants and without aggregating multiple responses per iteration (see \cite{harrison2020gsp, vanrijn2021exploring} for an exploration of these options). Another possible modification is changing the interface so that there are more sliders per interaction.  

Another approach to disentangle prosody could be to initially obtain the speaker embeddings from a speaker verification network and at a later stage to learn them with the other components of the model. This initialization with speaker verification embeddings can be a reasonable starting point. In the present study, we used the first ten principal components computed on the SpeakerNet Embeddings, but future research can explore alternative parametrizations of the speaker latent space that potentially better align with human perception and might explain more variance.

\section{Conclusion}
In this paper, we presented a human-in-the-loop approach for generating speech personalized to a specific face. To this end, we used an embedding space of a speaker verification network as input to a state-of-the-art TTS model. The embeddings were modified collaboratively by the use of a GSP experiment in order to find a voice that is consistent with shown images of realistic faces, portraits, and cartoon-style faces. Our evaluation showed that over the course of our experiment, the MOS of the generated speech, representing the consistency between the speech and the shown faces, increased from 2.7 to 4.0. This indicates that our approach is promising for synthesizing highly personalized speech for a specific speaker's visual appearance. 
Taken together our results open up a vast range of creative and practical applications including personalized voices in audiobooks and games, personalized speech assistants, and individualized voices for people with speech-impairments.

\section{Acknowledgments}
This work has partially been funded by the European Union Horizon 2020 research and innovation programme, grant agreement 856879 (PRESENT). 
\clearpage
\bibliographystyle{IEEEtran}

\bibliography{main}

\end{document}